\documentclass[twoside,twocolumn,9pt]{article}
\usepackage{extsizes}
\usepackage[super,sort&compress,comma]{natbib}
\usepackage[version=3]{packages/mhchem}
\usepackage[left=1.5cm, right=1.5cm, top=1.785cm, bottom=2.0cm]{geometry}
\usepackage{balance}
\usepackage{packages/widetext}
\usepackage{times,mathptmx}
\usepackage{packages/sectsty}
\usepackage{graphicx}
\usepackage{lastpage}
\usepackage[format=plain,justification=raggedright,singlelinecheck=false,font={stretch=1.125,small,sf},labelfont=bf,labelsep=space]{caption}
\usepackage{float}
\usepackage{fancyhdr}
\usepackage{fnpos}
\usepackage[english]{babel}
\usepackage{array}
\usepackage{packages/droidsans}
\usepackage{charter}
\usepackage[T1]{fontenc}
\usepackage[usenames,dvipsnames]{color}
\usepackage{color}
\usepackage{xcolor}
\setcitestyle{square}
\usepackage{setspace}
\usepackage[compact]{titlesec}
\usepackage{epstopdf}
\usepackage{multirow} 

\definecolor{cream}{RGB}{222,217,201}

\begin{document}

\pagestyle{fancy}
\thispagestyle{plain}
\fancypagestyle{plain}{

}

\makeFNbottom
\makeatletter
\renewcommand\LARGE{\@setfontsize\LARGE{15pt}{17}}
\renewcommand\Large{\@setfontsize\Large{12pt}{14}}
\renewcommand\large{\@setfontsize\large{10pt}{12}}
\renewcommand\footnotesize{\@setfontsize\footnotesize{7pt}{10}}
\makeatother

\renewcommand{\thefootnote}{\fnsymbol{footnote}}
\renewcommand\footnoterule{\vspace*{1pt}%
\color{cream}\hrule width 3.5in height 0.4pt \color{black}\vspace*{5pt}}
\setcounter{secnumdepth}{5}

\makeatletter
\renewcommand\@biblabel[1]{#1}
\renewcommand\@makefntext[1]%
{\noindent\makebox[0pt][r]{\@thefnmark\,}#1}
\makeatother
\renewcommand{\figurename}{\small{Fig.}~}
\sectionfont{\sffamily\Large}
\subsectionfont{\normalsize}
\subsubsectionfont{\bf}
\setstretch{1.125} 
\setlength{\skip\footins}{0.8cm}
\setlength{\footnotesep}{0.25cm}
\setlength{\jot}{10pt}
\titlespacing*{\section}{0pt}{4pt}{4pt}
\titlespacing*{\subsection}{0pt}{15pt}{1pt}
\fancyhead{}
\renewcommand{\headrulewidth}{0pt}
\renewcommand{\footrulewidth}{0pt}
\setlength{\arrayrulewidth}{1pt}
\setlength{\columnsep}{6.5mm}
\setlength\bibsep{1pt}

\makeatletter
\newlength{\figrulesep}
\setlength{\figrulesep}{0.5\textfloatsep}

\newcommand{\topfigrule}{\vspace*{-1pt}%
\noindent{\color{cream}\rule[-\figrulesep]{\columnwidth}{1.5pt}} }

\newcommand{\botfigrule}{\vspace*{-2pt}%
\noindent{\color{cream}\rule[\figrulesep]{\columnwidth}{1.5pt}} }

\newcommand{\dblfigrule}{\vspace*{-1pt}%
\noindent{\color{cream}\rule[-\figrulesep]{\textwidth}{1.5pt}} }

\makeatother
\twocolumn[
  \begin{@twocolumnfalse}
\vspace{1cm}
\sffamily
\begin{tabular}{m{18.0cm} p{0.0cm} }
	\noindent\Large{\textbf{{\color{olive}BN Doping in the Realm of Two-Dimensional Fullerene Network for Unparalleled Structural, Electronic, Optical, and HER  Advancements: A Cutting-Edge DFT Investigation}}
} \\
\\
\noindent\large{Vivek K. Yadav$^{a*}$} \\
\noindent\small{(a) Department of Chemistry, University of Allahabad, Prayagraj, UP. India 221001}\\
\\
\noindent\normalsize{\textit{{\color{blue} The doping of lighter non-metals like boron and nitrogen into graphene represents a promising advancement in the field of nanoelectronic devices, particularly in the development of field-effect transistors (FETs). These doped two-dimensional (2D) materials offer improved stability and enhanced adsorption characteristics compared to pure graphene. Notably, It displays semiconducting behavior, resulting in higher conductivity and carrier mobility. This study investigates the structural, electronic, optical, and conductivity/carrier transport properties of 2D polymer sheets made of fullerene, both with and without boron and nitrogen doping. We employ density functional theory (DFT) with PBE and HSE functionals, considering the inclusion of van der Waals (vdW) interactions. The research findings indicate that the 2D sheets of $C_{60}$, $C_{58}B_{1}N_{1}$, and $C_{54}B_{3}N_{3}$ exhibit band gaps of approximately 0.97 eV (1.5 eV), 1.08 eV (1.9 eV), and 1.05 eV (1.6 eV), respectively, as obtained from PBE (HSE) calculations. Moreover, according to the deformation potential theory, both doped sheets exhibit ultra-high conductivity ($\sim$ $10^{14}$ $ \Omega^{-1} cm^{-1} s^{-1}$ at elevated temperature). These results are promising and underscore the significance of a single pair of BN dopants in fullerene ($C_{58}B_{1}N_{1}$) monolayers for the advancement of next-generation 2D nano-electronic and photonics applications.}}}
\end{tabular}
 \end{@twocolumnfalse} \vspace{0.5cm}
  ]
\renewcommand*\rmdefault{bch}\normalfont\upshape
\rmfamily
\section*{}
\vspace{-1cm}
\footnotetext{\textit{$*$Corresponding author Email: vkyadav@allduniv.ac.in}}

\section{Introduction}
In recent decades, the allure of metal-free semiconductors with narrow bandgap has captured significant attention. These materials have proven to be exceptionally versatile for a myriad of electronic applications, including infrared devices, light-emitting diodes, electro-catalysts, and thermo-photovoltaics. 
In the realm of material science and engineering, the emergence of two-dimensional nanomaterials has opened an exciting new chapter, particularly following the groundbreaking discovery of graphene by a collaborative effort in the past decade \cite{novo,novo1,antio} and this new material (graphene), boasted with massless Dirac fermions and showcased extraordinary carrier mobility in the range of 1 to 4 $\times 10^5 cm^2 V^{-1}s^{-1}$\cite{fang}. These two-dimensional (2D) nanomaterials possess a remarkable volume-to-surface ratio, granting them exceptional properties that find diverse applications in various scientific and technological fields.This unique characteristic sets 2D nanomaterials apart from other forms of nanomaterials (0D, 1D, and 3D) as well as their bulk counterparts. It paves the way for a fresh and profound exploration of an abundance of novel 2D nanomaterials, igniting immense interest in utilizing them across multidisciplinary sectors for practical applications\cite{guorui,luo,zhu}. The versatility and potential of 2D nanomaterials promise to revolutionize numerous industries and drive innovation in the scientific community.\cite{milon}
However, one limitation of pristine graphene lies in its lack of a bandgap, which impedes certain applications. To overcome this hurdle, researchers turned to the enticing realm of doping. Introducing hetero-atoms like boron, nitrogen, or phosphorus into graphene gives rise to a fascinating class of materials with a finite bandgap, while still maintaining carrier mobility comparable to graphene\cite{vivek1}.
The direct creation of 2D materials from graphene by incorporating lighter elements such as boron, nitrogen, and phosphorus has garnered tremendous interest, primarily due to their extraordinary electronic and opto-electronic potential\cite{vivek2,cnrrao}. The strategic selection of foreign elements in graphene, ensuring an optimal carbon to doping element ratio, plays a pivotal role in tailoring the electronic properties of these newly crafted 2D materials, thereby enhancing their applicability in cutting-edge devices\cite{cnr1,cnr2,cnr3}.
\\
In the quest to fine-tune the properties of nanomaterials, several specific methods have emerged, including surface defects\cite{omid,sun}, metal decoration\cite{nemati, esra, lins}, and transition metal doping\cite{tizr, mogh}. These techniques are widely explored both through experimental studies and density functional theory (DFT)-based calculations. In certain cases, non-metal elements such as Boron and  Nitrogen also play a crucial role in enhancing surface reactivity\cite{cnr2}. The wealth of comprehensive information gathered from these investigations highlights the widespread and efficient nature of nanomaterials' tuning properties in the realms of material physics and engineering. This knowledge opens up exciting possibilities for tailoring nanomaterials to meet the specific requirements of diverse applications in various scientific and engineering fields\cite{wang,wang1,maroto,laurent}.\\
In their research, Peng and colleagues discovered that using a weakly screened hybrid functional in conjunction with time-dependent Hartree-Fock calculations, to incorporate the exciton binding energy, successfully replicated the experimentally measured optical band gap of monolayer $C_{60}$\cite{peng}. The different phases of monolayer fullerene networks were found to possess suitable band gaps, exhibiting high carrier mobility and appropriate band edges, making them thermodynamically favorable for driving overall water splitting. Furthermore, the team explored the optical properties of monolayer $C_{60}$, unveiling that the various phases of fullerene networks displayed distinct absorption and recombination behaviors. This characteristic grants them unique advantages, serving as either efficient electron acceptors or electron donors in the domain of photocatalysis. These findings highlight the immense potential of monolayer fullerene networks for applications in the field of energy conversion and utilization.
\\
The unique cage structure of $C_{60}$ enables it to exhibit remarkable quantum efficiency in photocatalytic reactions. This is due to its large surface area, abundant micropores, increased surface active sites, and efficient electron transport properties. In the realm of photocatalysis, $C_{60}$ can enhance the photocatalytic activity through various mechanisms. It can function as an electron acceptor, promoting rapid carrier separation, or act as an energy transfer mediator. Additionally, $C_{60}$ can also serve as an electron donor due to its high photosensitivity\cite{hou}.
When incorporated into composite materials, fullerene enhances their crystallization by reducing defects and improves the overall stability of the composites, thereby further boosting their photocatalytic efficiency. Intriguingly, $C_{60}$ itself shows great promise as a hydrogen storage material, making photocatalytic water splitting using fullerene a convenient approach for simultaneous hydrogen production and storage. These exceptional properties make $C_{60}$ a highly versatile and valuable component in various photocatalytic applications with significant potential for sustainable energy generation\cite{zhao,yoon,wang2,wang3,durbin}.
A 2D material with covalently bonded fullerene network structures has been synthesized, yielding two configurations: a few-layer quasi-tetragonal phase (qTP) and a monolayer quasi-hexagonal phase (qHP)\cite{hou}. Monolayer $C_{60}$ stands out among other 2D materials\cite{kim,zhang1,yu,pier,hill} due to its larger surface area and increased active sites resulting from the quasi-0D network structures of $C_{60}$ cages. Moreover, it exhibits excellent thermodynamic stability and high carrier mobility\cite{hou}, making it a promising candidate for photocatalytic water splitting. However, all theoretical calculations currently underestimate the band gap of monolayer $C_{60}$ by at least 10\% \cite{tromer, yul,yuan}. A precise understanding of the band structures is crucial for exploring band edge positions for water splitting and optical absorption for photocatalysis.
Recently, a significant advancement was made in the field of 2D carbon materials with the successful creation of a single-crystal monolayer quasi-hexagonal-phase fullerene ($C_{60}$)\cite{hou}. This remarkable material possesses a semiconducting bandgap of approximately 1.6 eV, a breakthrough that addresses the challenge of null bandgaps observed in other 2D carbon-based materials\cite{zhang2}. The process of obtaining the monolayer polymeric $C_{60}$ involved an organic cation slicing strategy, enabling the exfoliation of quasi-hexagonal bulk single crystals and yielding $C_{60}$ 2D crystals of substantial size through an interlayer bonding cleavage approach. In the crystal structure of these materials, covalently bonded cluster cages of $C_{60}$ form in a plane, resulting in two stable phases: closely packed quasi-hexagonal (qHPC$C_{60}$) and quasi-tetragonal (qTP$C_{60}$). These phases, qHP$C_{60}$ and qTP$C_{60}$, exhibit remarkable crystallinity and exceptional thermodynamic stability. Their moderate bandgap and unique topological structure make them highly promising for potential applications in nanoelectronics.
Despite the valuable insights provided by the original work \cite{zhang2} regarding the physical properties of these 2D $C_{60}$ crystals, a comprehensive description of their electronic, optical, and mechanical characteristics is still awaited. Further dedicated research and exploration in these areas offer significant potential to unlock the full range of their capabilities.
According to Gao and colleagues' first-principles findings, they have demonstrated that an $O_{2}$ molecule can easily adsorb and undergo partial reduction on the N-C complex sites (Pauling sites) of N-$C_{60}$, without encountering any activation barrier. Subsequently, the partially reduced $O_{2}$ can directly react with $H^{+}$ and additional electrons, leading to the completion of the water formation reaction (WFR) without any activation energy barrier. As a result, N-$C_{60}$ fullerene has the potential to serve as a promising cathode catalyst for hydrogen fuel cells\cite{gao1}.
\\
In this article, we present a comprehensive exploration of the properties of $C_{60}$, $C_{58}B_{1}N_{1}$, and $C_{54}B_{3}N_{3}$ nanosheets. To achieve this, we outline the calculation methodology by detailing the employed DFT methods. Subsequently, in the Results and Discussion section, we delve into the structural and lattice dynamic stability of both the doped materials. Moving forward, we conduct an in-depth investigation of the electronic band structure and projected density of states of the nanosheets. Additionally, we analyze the mechanical and optical properties of proposed sheets, and we compare the results with polymeric fulurene. Finally, we conclude with a succinct summary of our significant findings obtained throughout this study.
\section{Methodology}
Utilizing the density functional theory (DFT) framework, first-principle simulations provide noteworthy quantum insights into nanostructured materials. This approach offers high precision and cost-effectiveness\cite{maroto,laurent}.
In this study, we performed first-principle calculations employing the QUANTUM ESPRESSO\cite{giannozzi2009quantum} software package, utilizing the density functional theory (DFT) framework. The Perdew-Burke-Ernzerhof\cite{perdew1996generalized} functional was employed to describe the electron-ion interaction, adopting the generalized gradient approximation (GGA)\cite{perdew1992atoms}. To achieve the optimized structural configuration of fullerenes-based systems, we utilized a plane wave basis set with a kinetic energy cut-off of 50 Ry for electron density and 500 Ry for charge density. The Brillouin zone (BZ) integration was conducted using a uniform Monkhorst-Pack\cite{monkhorst1976special} k-point grid of $10\times10\times1$ for geometry optimization and $20\times20\times1$ for electronic structure calculations. To ensure accurate results, we fully relaxed the atomic positions and cell parameters until an energy convergence of $10^{-8}$ eV was achieved. To account for van der Waals interactions, we incorporated DFT-D3 dispersion corrections\cite{grimme2010consistent}. Given that the PBE functional tends to underestimate bandgap values, we additionally employed the HSE functional\cite{hse} to compute the bandgap with k-points set to $10\times10\times1$. For our investigations, the unit cell was appropriately designed to include one and three pairs of BN atoms doped in $C_{60}$ fullerene. To prevent any unwanted interactions between periodic images, a large vacuum region of 40\AA{} was introduced in the z-direction perpendicular to the sheet.\\
\section{Results and discussion}
\subsection{Structural properties}
\begin{figure}
	\centering
	\includegraphics[width=3.5in]{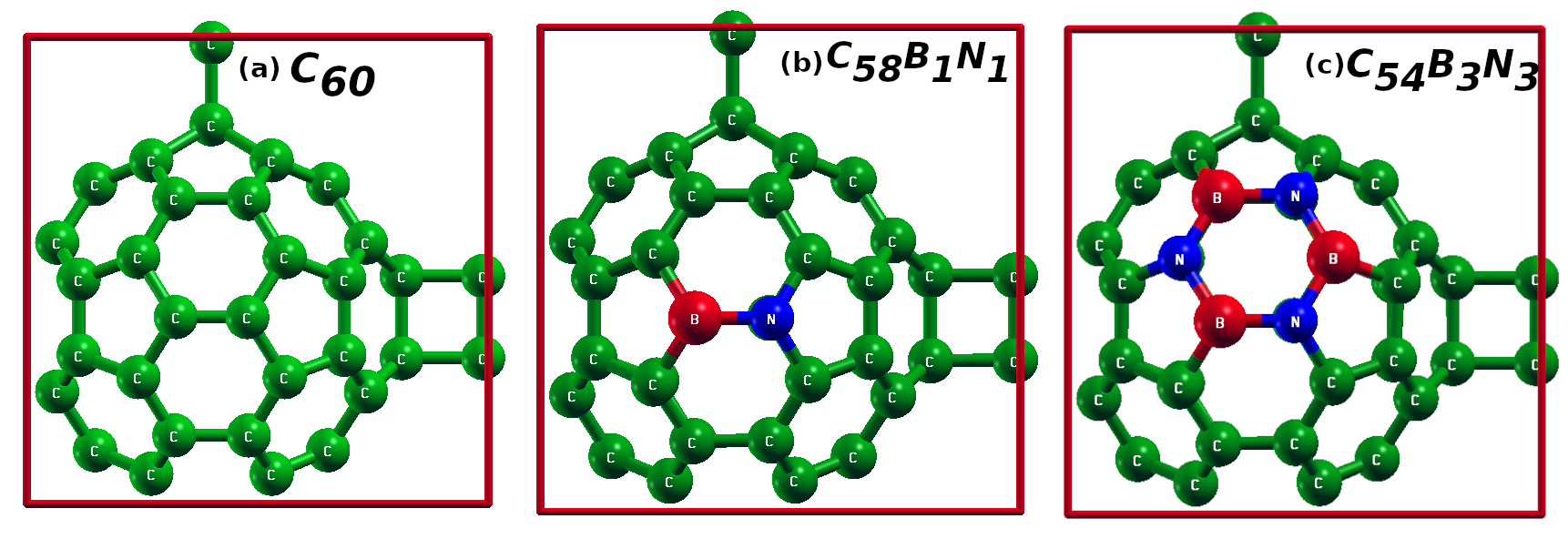}
	\caption{(a,b,c) Schematic representation of the optimized atomic configuration of (a) $C_{60}$, (b) $C_{58}B_{1}N_{1}$, and (c) $C_{54}B_{3}N_{3}$, unit cell. The green, red, and blue ball denotes the Carbon, Boron and Nitrogen atoms, respectively.}
	\label{Fig:fig1}
\end{figure}
The optimized configurations of $C_{60}$, $C_{58}B_{1}N_{1}$, and $C_{54}B_{3}N_{3}$ are illustrated in Figure~\ref{Fig:fig1}(a,b,c). Both atomic positions and cell parameters underwent relaxation during the optimization process. The calculated lattice parameters "a" and "b" for $C_{60}$, $C_{58}B_{1}N_{1}$, and $C_{54}B_{3}N_{3}$ along the x-axis (y-axis) were found to be 9.00074 \AA (9.10137 \AA), 9.01298 \AA (9.09549 \AA) and, 9.00580 \AA (9.11534 \AA), respectively. Since a B-N pair mimics a C-C bond in terms of electron count, the overall cell volume remains consistent with that of $C_{60}$ following B-N doping.\\
The thermodynamic and structural stability of the three proposed nanosheets were assessed by calculating its formation and cohesive energies. The formation energy ($E_{for}$) is computed using equation~\ref{eq:for}.
\begin{equation}\label{eq:for}
	E_{for} = [E_{system}-(n_{CC} \mu_{CC} + n_{BN} \mu_{BN}]/n 
\end{equation}
whereas their cohesive energy ($E_{coh}$) can be calculated by using equation~\ref{eq:Coh}
\begin{equation}\label{eq:Coh}
E_{coh} = [E_{tot}-\sum_{i}^{}n_iE_i]/n \quad (i={\textnormal {C, B, N}})
\end{equation}
Here in equation ~\ref{eq:for}, the $n_{CC}$ and $n_{BN}$ denote the CC and BN pairs in the proposed sheets and the total number of C-C bond into prestine $C_{60}$ is taken to be 60. The symbol $\mu_{CC}$ and $\mu_{BN}$ corresponds to the chemical potential of C–C and B–N, respectively. The chemical potential $\mu_{CC}$ and $\mu_{BN}$ obtained from graphene and BN sheet were -310.150 and -358.352 eV, respectively. The $E_{for}$ for $C_{60}$, $C_{58}B_{1}N_{1}$, and $C_{54}B_{3}N_{3}$ is found to be 147.834, 148.038, and 148.412, respectvely.
In the equation ~\ref{eq:Coh}, $E_{tot}$  denotes the total energy of three individual sheets, $E_i$ represents the gas phase atomic energies of Carbon, Boron, and Nitrogen and $n$ represents the total number of atoms in the sheet. The calculated cohesive energies of $C_{60}$, $C_{58}B_{1}N_{1}$, and $C_{54}B_{3}N_{3}$ were found to be -8.757, -8.717 and -8.672, in units of eV, respectively. This shows that stabilty decreases monotonically with increasing B-N dopant concentration. A molecular structure with a more negative cohesive energy value indicates greater structural stability\cite{rad}. In our observations, it is evident that pristine $C_{60}$ exhibits a notably higher negative cohesion energy value, whereas doped nanosheets showcase predominant variations in other properties.
Further it is recomended to evaluate the lattice dynamical stability of the nanosheets via the phonon dispersion calculation, but due to bigger unit cell size and limited computational resources, we are not performing it in current work. It has been confirmed in the literature that the $C_{60}$ polymer exhibits positive frequencies across the Brillouin zone, indicating dynamical stability\cite{peng}. 
\subsection{Chemical reactivity}
The introduction of dopants into a material brings about a notable modification in its chemical reactivity. Consequently, the computation of comprehensive descriptive-DFT parameters, such as chemical potential $\mu$\cite{parr1}, overall softness S, overall hardness $\eta$\cite{parr2}, and electrophilicity $\omega$\cite{niazi}, offers an effective approach to understanding the impact of doping on chemical reactivity. The equations are as follows:
\begin{equation}\label{eq:chem}
\begin{aligned}
                 \mu = \frac{E_{HOMO}+E_{LUMO}}{2}  \\
                \eta = \frac{E_{LUMO}-E_{HOMO}}{2}  \\
                   S = \frac{1}{2\eta}  \\
              \omega = \frac{\mu^{2}}{2\eta}
\end{aligned}
\end{equation}\\
The DFT calculations and the aforementioned global indices, computed using the equations ~\ref{eq:chem} provided, are presented in Table 1. The global chemical potential of the pristine $C_{60}$ nanosheet is determined to be -1.083 eV, which is comparatively lower than that of the other two Co-doped $C_{60}$-BN nanosheets. This indicates that the chemical activity of the pristine nanosheet is more pronounced than the other two. To address any uncertainties concerning the BN-doped $C_{60}$ nanosheet, the maximum hardness (where softness is the reciprocal of hardness) and the minimum electrophilicity principle can provide clarity. As depicted in Table 1, the pristine $C_{58}B_{1}N_{1}$ nanosheet exhibits the highest hardness (0.546 eV) and the lesser electrophilicity (0.891 eV). In summary, our findings lead us to conclude that the chemical reactivity of the pristine $C_{60}$ nanosheet surpasses that of the two doped BN nanosheets ($C_{58}B_{1}N_{1}$ and $C_{54}B_{3}N_{3}$). The reactivity order can be described as follows: $C_{58}B_{1}N_{1}$ > $C_{54}B_{3}N_{3}$ > $C_{60}$. Hence, it can be asserted that the introduction of dopants has expanded the potential applications of these nanosheets, enabling their utilization in fields like battery and medical devices.
\begin{table*}
\begin{centering}
\begin{tabular}{ |p{4.00cm}||p{3.00cm}|p{3.00cm}|p{3.00cm}|  }
 \hline
	\multicolumn{4}{|c|}{{\bf Systems}} \\
 \hline
{\bf Properties}& $C_{60}$ &$C_{58}B_{1}N_{1}$&$C_{54}B_{3}N_{3}$\\ \hline \hline
Lattice parameter {\bf a}(\AA) & 9.00074 &9.01298&9.00580\\  \hline
Lattice parameter {\bf b}(\AA) & 9.10137 &9.09549&9.11534\\  \hline
Cohesive Energy {\bf $E_{coh}$}(eV) & -8.757  &-8.717& -8.672\\  \hline
HOMO (eV) & -1.570    &-1.533&-1.435\\  \hline
LUMO (eV) & -0.597    &-0.441&-0.387\\  \hline
$E_{gap}$ (eV) & 0.97 &1.09& 1.05\\  \hline
$E_{fermi}$ (eV) & -1.076    &-0.991&-0.919\\  \hline
Chemical Potential {\bf $\mu$}(eV) &-1.083 &-0.987&-0.911\\  \hline
Hardness {\bf $\eta$} (eV) &0.487 &0.546 &0.524\\  \hline
Softness {\bf S}($eV^{-1}$) &1.027 &0.915 &0.954\\  \hline
Electrophilicity {\bf $\omega$} (eV) & 1.206 &0.891&0.792\\  \hline
Elastic Constant {\bf $C_{2D}$} ($Jm^{-2}$) & 384.748  &379.493&371.838\\  \hline
Conductivity {\bf $\sigma/\tau$} at 300K & $1.91$x$10^{10}$ & $3.70$x$10^{14}$ & 0\\  \hline
Work function (eV) & 4.854  &4.744  &4.641\\  \hline
\end{tabular}
\caption{Lattice parameter (\AA), Cohesion energy $E_{coh}$ (eV), HOMO energy (eV), Fermi Level (eV), LUMO energy (eV), Band gap $E_{gap}$ (eV), Chemical Potential $\mu$ (eV), Hardness $\eta$ (eV), Softness S (eV), Electrophilicity $\omega$ (eV), Elastic Constant ($Jm^{-2}$), Conductivity and work function (eV) for (a) $C_{60}$, (b) $C_{58}B_{1}N_{1}$, and (c) $C_{54}B_{3}N_{3}$, respectively.}\label{Fig:bandgap_block}
\end{centering}
\end{table*}

\subsection{Electronic Structure}
\begin{figure}
  	\centering
  	\includegraphics[width=3.5in, height = 2.5in]{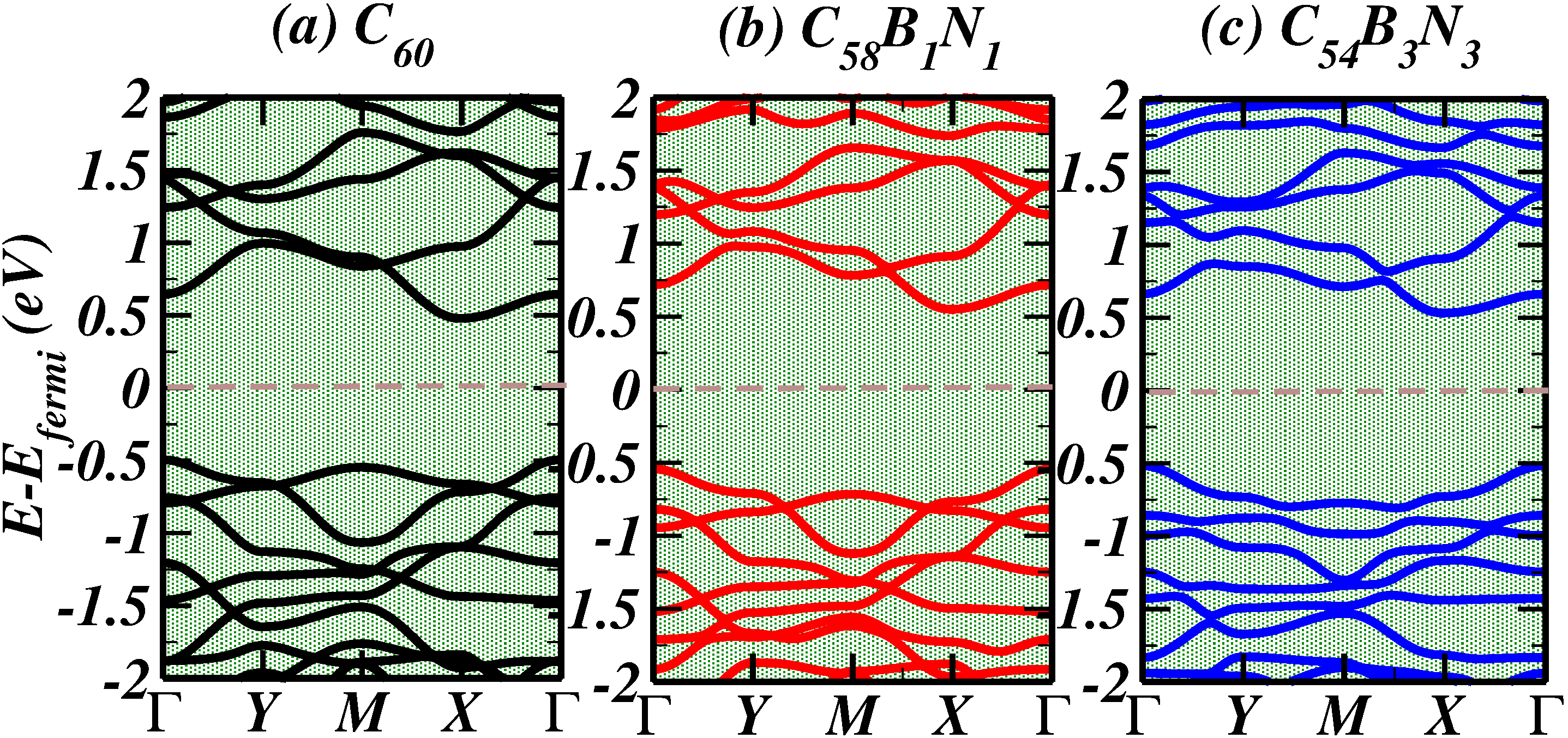}
  	\caption{The calculated band structures of (a) $C_{60}$, (b) $C_{58}B_{1}N_{1}$, and (c) $C_{54}B_{3}N_{3}$}\label{Fig:Band_str}
  \end{figure}

\begin{figure}
	\centering
	\includegraphics[width=3.0in]{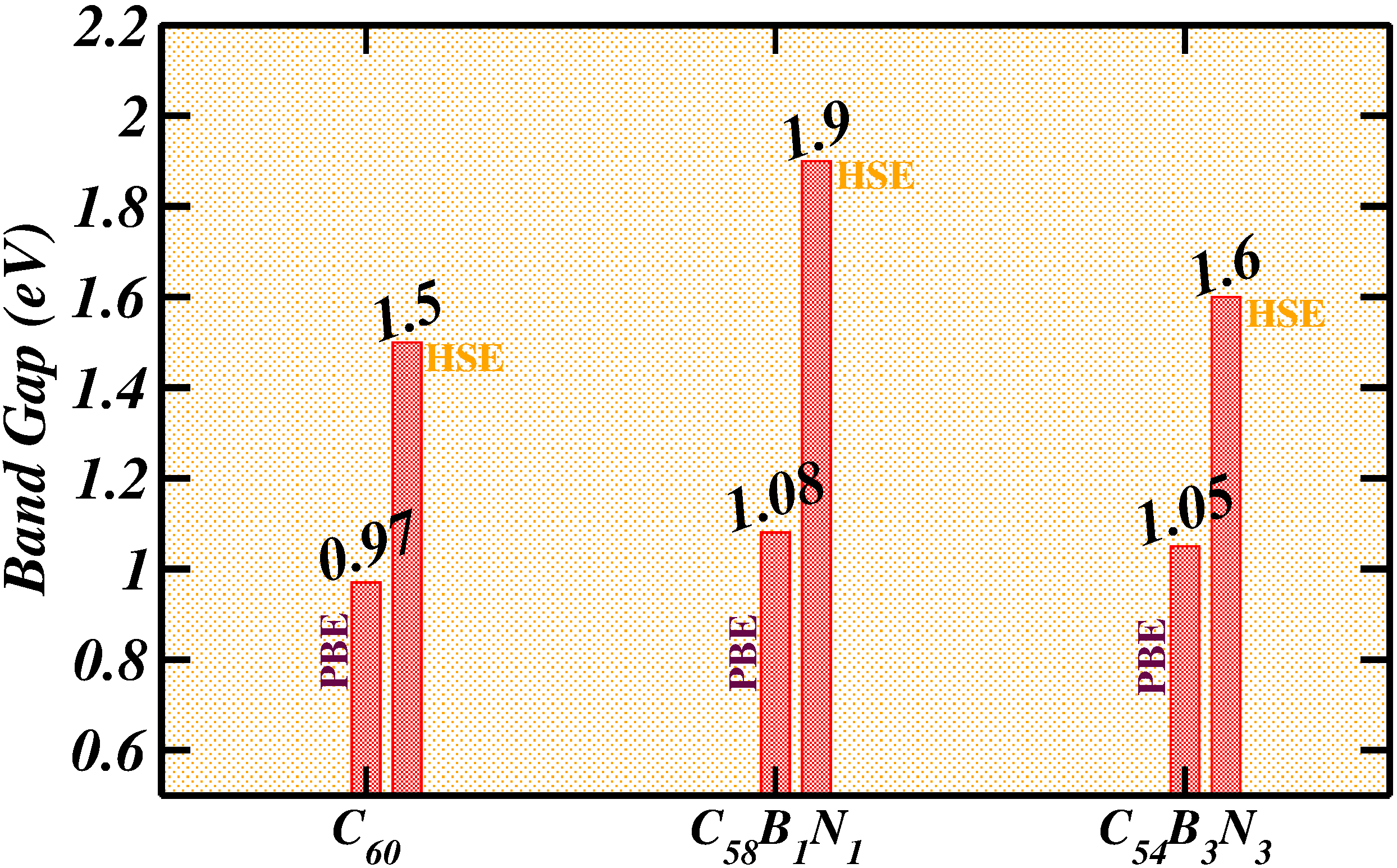}
	\caption{Band gap of (a) $C_{60}$, (b) $C_{58}B_{1}N_{1}$, and (c) $C_{54}B_{3}N_{3}$ from PBE and HSE functional, respectively.}
	\label{Fig:bg}
\end{figure}

\begin{figure}
  	\centering
  	\includegraphics[width=3.5in, height = 5.0in]{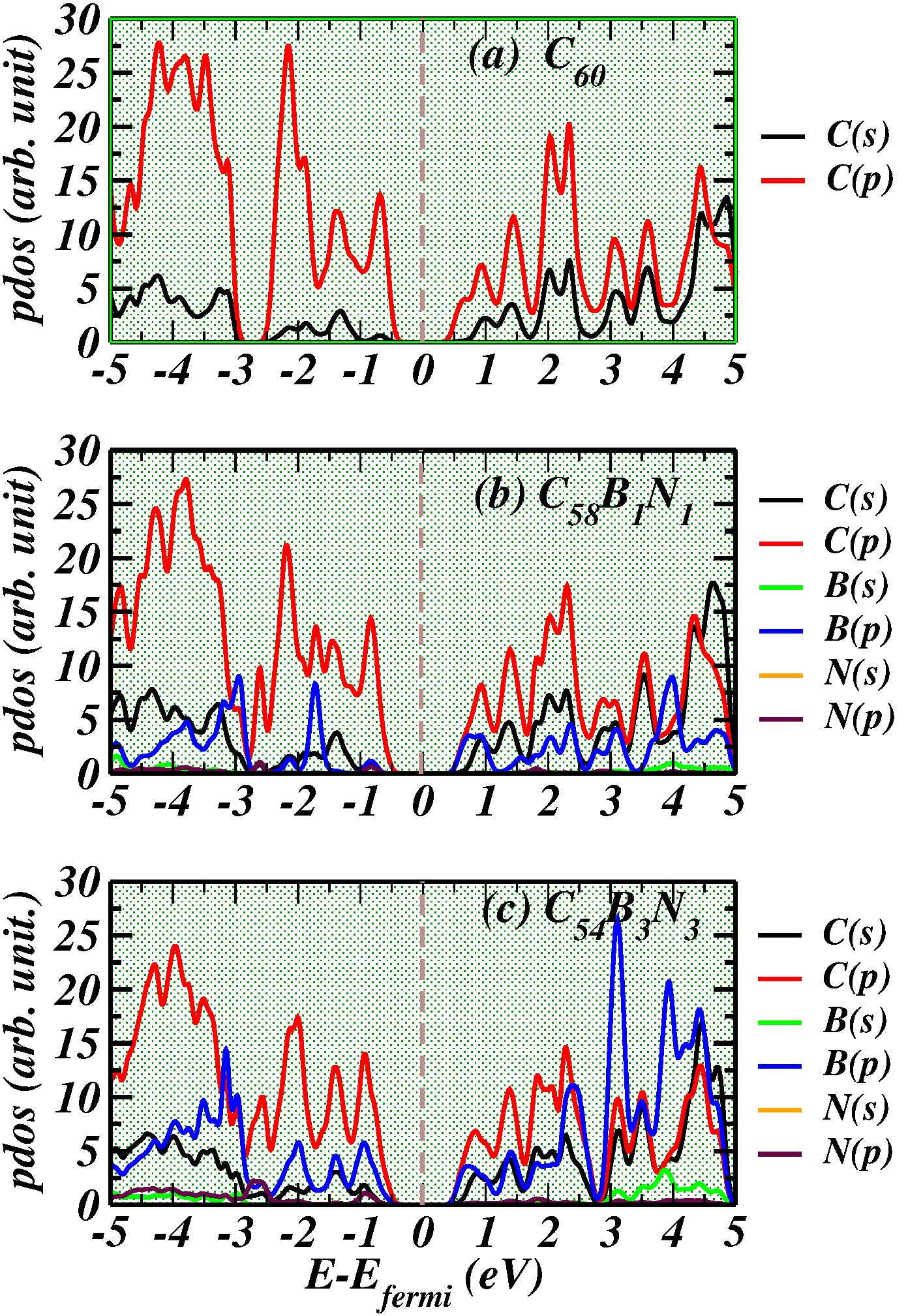}
  	\caption{Total density of states and partial density of states of (a) $C_{60}$, (b) $C_{58}B_{1}N_{1}$, and (c) $C_{54}B_{3}N_{3}$.}
	  \label{Fig:pdos}
  \end{figure}

\begin{figure}
	\centering
	\includegraphics[width=3.5in]{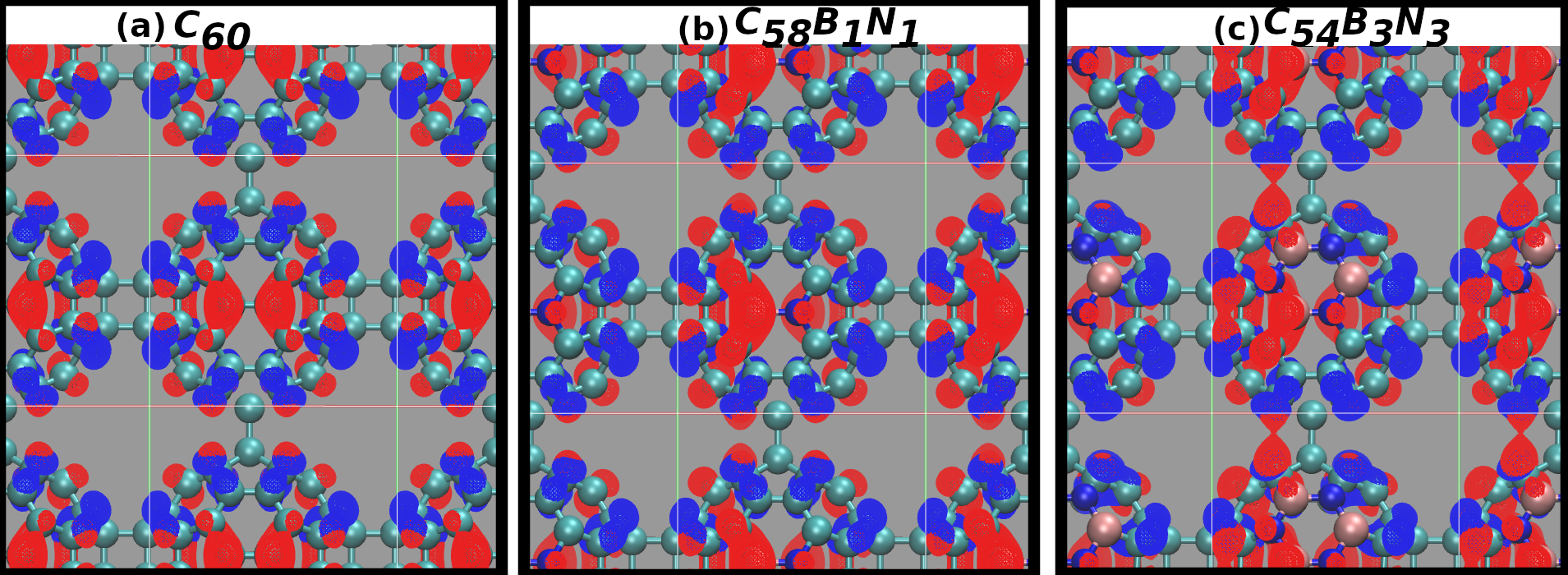}
	\caption{HOMO and LUMO plot using PBE functional for (a) $C_{60}$, (b) $C_{58}B_{1}N_{1}$, and (c) $C_{54}B_{3}N_{3}$, respectively. The isosurface value is set at 0.009 e\AA$^{-3}$}
	\label{Fig:hl}
\end{figure}

The calculated band structure of (a) $C_{60}$, (b) $C_{58}B_{1}N_{1}$, and (c) $C_{54}B_{3}N_{3}$ are shown in Figure~\ref{Fig:Band_str}. From the band structure, it is seen that all the three systems exhibit an indirect bandgap with valence band maximum (VBM) located at $\Gamma$-point and conduction band minimum (CBM) occurs at point-X of the Brillouin zone. It is clear from the Figure~\ref{Fig:Band_str} that all the three system exhibit substantial dispersion in the band structures. The structures show parabolic nature of CBM and VBM which implies that deformation potential theory can be applied for such systems. Moreover, as it is well-known fact that DFT with PBE functional always underestimates the bandgap\cite{perdew,burke}. Therefore, we additionally employed the HSE functional to accurately determine the bandgap. The calculated bandgap for (a) $C_{60}$,(b) $C_{58}B_{1}N_{1}$, and (c) $C_{54}B_{3}N_{3}$ were 0.97 (1.5), 1.08 (1.9) and 1.05 (1.6) in units of eV, by employing PBE (HSE) functional, respectively as shown in Figure~\ref{Fig:bg}. 
From HSE results, one can predict that B $C_{58}B_{1}N_{1}$ is a wide bandgap semiconductor compared to  $C_{60}$ and $C_{54}B_{3}N_{3}$. It is well known that Silicon which has revolutionized the micro-electronic industry exhibit an indirect bandgap ($\sim$1.50 eV). Herein, we showed that both the BN doped $C_{60}$ sheets exhibits as an interesting material that shows a range of bandgap which is very close to that of Silicon. We conclude on based on bandgap that these doped material can be used for future nano- and opto-electronic devices (as band gap is below 2.0 eV). 
We also examined the total and partial density of states (PDOS) of all three systems, which are shown in Figure~\ref{Fig:pdos}. From the total density of states, it is apparent that $C_{58}B_{1}N_{1}$ larger bandgap than other two systems. From PDOS of$C_{58}B_{1}N_{1}$, it is observed that both CBM and VBM are dominated by Carbon p-orbital in all the three cases. This can also be justfy as percentage of Boron and Nitrogen is very less compared to carbon content into the system. In similar fashion we have computed and displays the HOMO and LUMO plot in Figure~\ref{Fig:hl} to elucidate the electronic arrangements for the $C_{60}$,$C_{58}B_{1}N_{1}$, and $C_{54}B_{3}N_{3}$ with isosurface value of 0.009 e\AA$^{-3}$. In all the systems, the red and blue mesh shows the HOMO and LUMO, respectively. It is concluded that the charge accumulation or depletion occour mostly around the BN site in the system, which makes the place ideal for adsorption and catlaysis\cite{vivek2}.
\subsection{Mechenical and optical properties}
For calculation of mechanical property like elastic constant, the E$^i = \Delta V_i / \varepsilon$ is computed where $\Delta E_i$ is the energy change of the i$^{th}$ band under cell compression or dilatation, $\varepsilon$ is the applied strain that is defined as $\varepsilon = \Delta l/l_0$, where $l_0$ is the equilibrium lattice constant in the transport direction and $\Delta l_0$ is the deformation of $l_o$.
The elastic modulus C$_{2D}$ of the longitudinal strain in the propagation directions (both x and y) of the longitudinal acoustic wave is calculated from parabolic fitting of the equation $(E - E_0 )/S_0 =C_{2D} \varepsilon^2 /2$, where $E$ denotes the total energy of deformed system, $E_0$ is the energy of system in equilibrium, and S$_0$ is the area of a system at equilibrium.  The strain is appled in both x anf y direction ranging from -2\% to +2\%. The computed values of elastic constant ($C_{2D}$) are 384.748, 379.493, and 371.838 for $C_{60}$, (b) $C_{58}B_{1}N_{1}$, and (c) $C_{54}B_{3}N_{3}$, in units of $Jm^{-2}$, respectively. The experimental measured elastic constant ($C^{\beta}$) for BN and graphene is 260, and 350 $Jm^{-2}$, respectively, which shows their strain induced stability\cite{lopez1,falin,pengq}.\\
Further, the conductivity of $C_{60}$, $C_{58}B_{1}N_{1}$ and $C_{54}B_{3}N_{3}$ were determined by means of the semiclassical Boltzmann theory within constant relaxation time approximation using BoltzTrap code\cite{BoltzTrap_code}. In this approach, the conductivity is calculated using the equation
\begin{equation}
\sigma_{\alpha,\beta}(T, \nu) = \Sigma_i \int \frac{dk}{8\pi}[-\frac{\partial f (T,\nu)}{\partial \varepsilon}] \sigma_{\alpha,\beta}(i,k)
\end{equation}
  \begin{figure}
  	\centering
  	\includegraphics[width=3.5in]{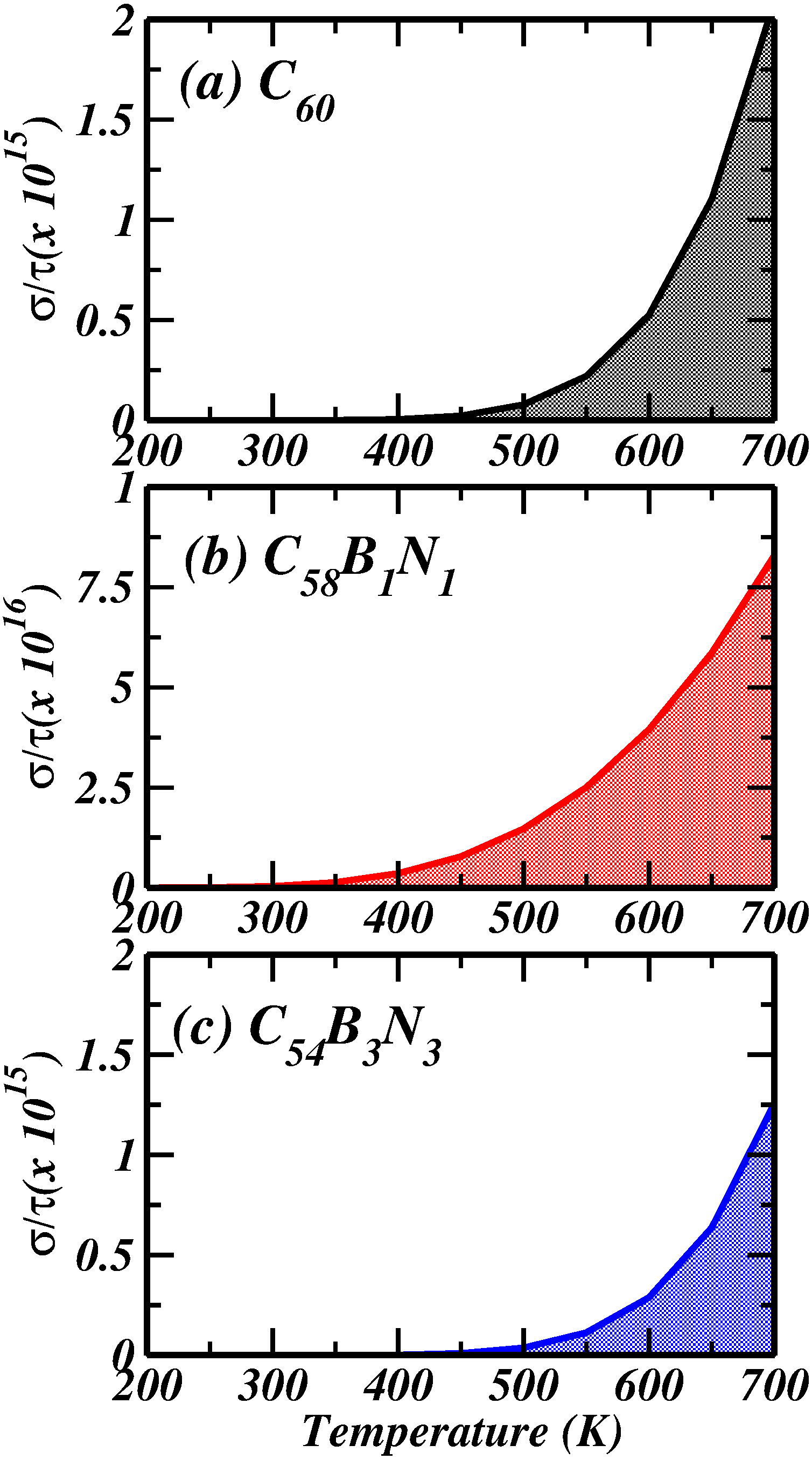}
	  \caption{Conductivity in units of ($ \Omega^{-1} cm^{-1} s^{-1}$) as a function of temperature, (a)$C_{60}$, (b) $C_{58}B_{1}N_{1}$, and (c) $C_{54}B_{3}N_{3}$,respectively.}\label{Fig:cond}
  \end{figure}
where $f$ is the Fermi-Dirac distribution function and $\nu$ is the chemical potential that is determined by the number of free carriers. Here, $\sigma_{\alpha,\beta}(i, k) = e^2 \tau_{i,k}v_\alpha(i,k)v_\beta(i,k)$ is the conductivity tensor in which $v(i,k) = \hbar^{-1} \frac{\partial \varepsilon_{i,k}}{\partial k_\alpha}$ represents the group velocity of $i^{th}$ band for $\alpha$ component. The calculated conductivity ($\sigma/\tau$) using BoltzTrap code is shown in Figure~\ref{Fig:cond}. From the figure, it is apparent that conductivity of $C_{58}B_{1}N_{1}$ increase with temperature much earlier as compared to $C_{60}$ and $C_{54}B_{3}N_{3}$. The conductivity of $C_{60}$, $C_{58}B_{1}N_{1}$ and $C_{54}B_{3}N_{3}$ at 300 K was found to be $1.91\times10^{10}$, $3.70\times10^{14}$, and $0$, in units of $ \Omega^{-1} cm^{-1} s^{-1}$, respectively. The presence of one pair of BN into $C_{60}$ enhances the conductivity by ten thousand times ($\sim$ 10,000) whereas when there pairs of BN is arranged in the $C_{60}$ in cyclic order then it make the system zero conductive at room temperature. Although our theoretical calculation shows that as we increase the room temperature by 50 K (i.e. above 350 K), the system $C_{54}B_{3}N_{3}$ also starts showing the conductive behavior and at very high temperature (around 750 K) it behaves similar to $C_{60}$ sheet. \\ 
The calculation of work function ($\Phi$) was performed using equation ~\ref{eq:wf}, and it represents the energy needed to extract an electron from the Fermi level to the vacuum.
\begin{equation}\label{eq:wf}
	\Phi = E_{vaccum}- E_{fermi}
\end{equation}
where $E_{vaccum}$ represents the vacuum potential and $E_{fermi}$ is the Fermi energy.
The computed work function of the pristine $C_{60}$ is 4.85 eV, demonstrating good congruence with the experimental value of 4.56 eV\cite{fujimoto,ryan} for graphene. Upon substitutional boron doping (SB), the work function increases, indicating a p-type characteristic. Conversely, in the case of substitutional nitrogen (SN) defects, the work function decreases aligning with its n-type doping nature. The work function for $C_{58}B_{1}N_{1}$ and $C_{54}B_{3}N_{3}$ is found to be 4.74 eV and 4.64 eV, respectively as shown in Figure ~\ref{Fig:wf}.
\begin{figure*}[ht]
\centering
        \includegraphics[width=6.0in, height = 2.0in]{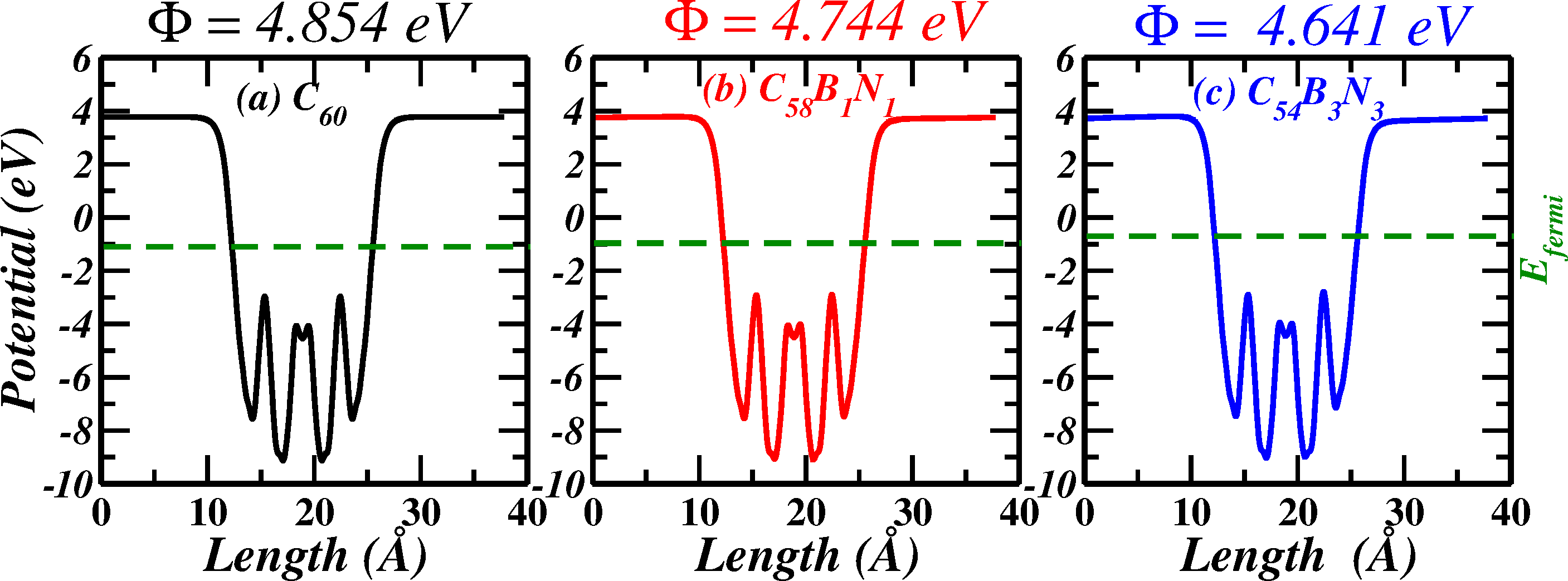}
        \caption{Work function of the (a) $C_{60}$, (b) $C_{58}B_{1}N_{1}$, and (c) $C_{54}B_{3}N_{3}$.}\label{Fig:wf}
\end{figure*}
 Notably, the work function of impurity-doped graphene exhibits variation based on the concentration of the impurity defects\cite{fujimoto}.\\
Hydrogen production through water splitting serves as a clean and environmentally friendly energy source\cite{vivek2,vivek_bcn}. The essential requirement for the hydrogen evolution reaction (HER) is that the semiconductor's band structure must straddle the redox potential of water\cite{showkat}. In order to substantiate this assertion, a comparison is made between the band edge positions of unblemished/BN doped $C_{60}$ sheets and the reduction potential of water $(H^{+}/H_{2})$. The positions of the valence and conduction bands' edges were determined by aligning them relative to the vacuum potential, as computed from the respective sheet calculations, using the equation $E^{VBE/CBE} = E^{VBM/CBM} - E_{vac}$.
In all three configurations of $C_{60}$ investigated in this study, the Fermi energy ($E_{fermi}$) is approximately -1.00 eV, resulting in an overpotential of around 3.75 eV relative to the HER redox potential. Figure~\ref{Fig:her} vividly demonstrates that the HER potential of water lies within the bandgap of the all the three configurations. This underscores the capability of $C_{60}$, $C_{58}B_{1}N_{1}$, and $C_{54}B_{3}N_{3}$ sheets to reduce $H^{+} to H_{2}$, thereby opening new avenues in the realm of photocatalysis. The valence and conduction band edges of these three systems effectively straddle the HER redox potential, setting them apart from the others. Consequently, we deduce that the all the three systems in Figure~\ref{Fig:her} hold promise as metal-free electrocatalysts. Moreover, the conduction and valence bands of $C_{58}B_{1}N_{1}$ (see Fig.2) obtained from calculations based on the PBE functional also exhibit alignment with the HER redox potential. 
\begin{figure}
  	\centering
  	\includegraphics[width=3in]{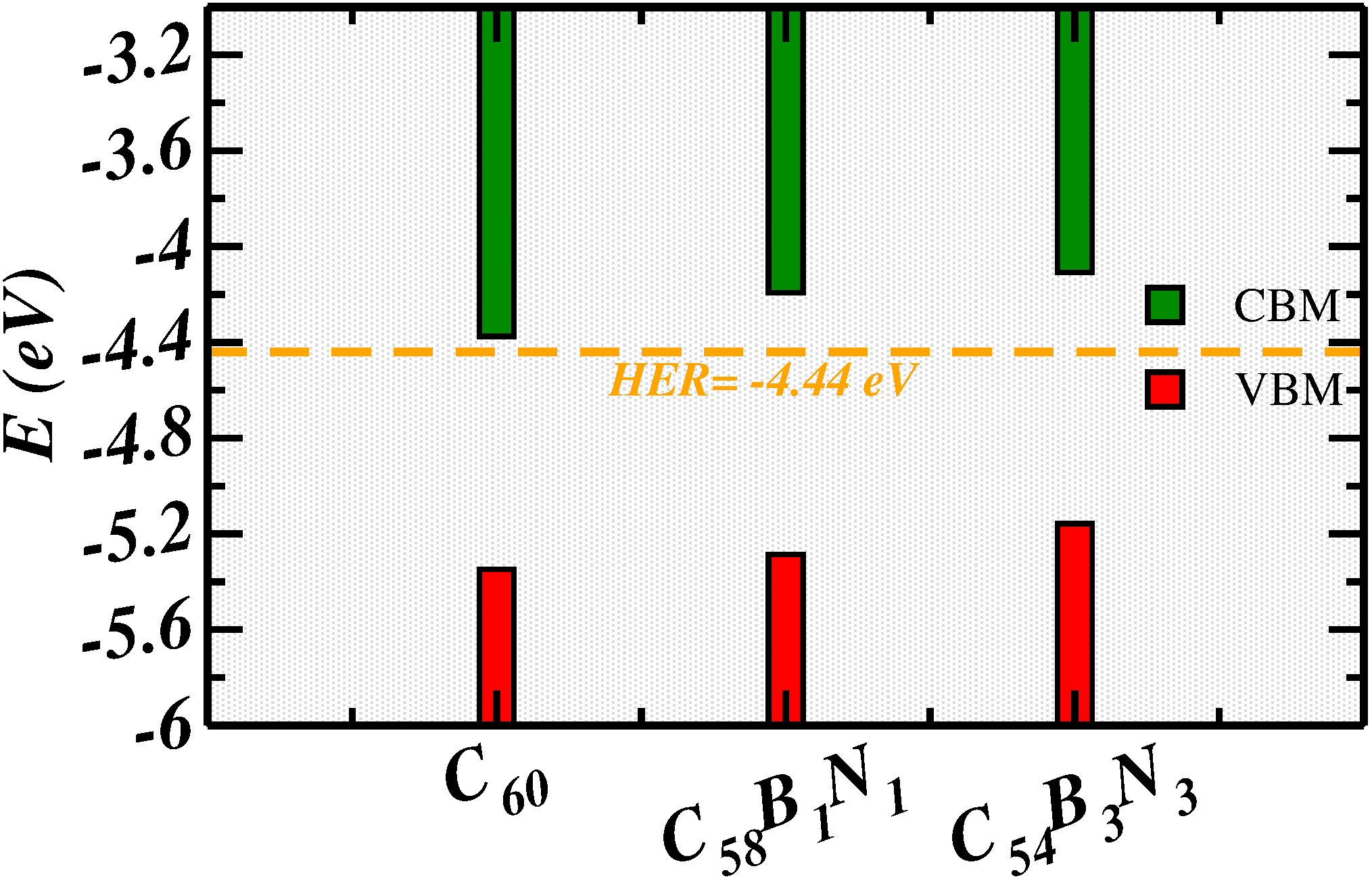}
	  \caption{Band-edge energies of valence (red) and conduction (green) band edges of the $C_{60}$, $C_{58}B_{1}N_{1}$, and $C_{54}B_{3}N_{3}$ conﬁgurations. The energy level for the hydrogen evolution reaction (HER) is denoted by a dashed line.}\label{Fig:her}
  \end{figure}
In conclusion, the outcomes of calculations grounded in the PBE functional indicate that the electronic structure of all the three systems under consideration here align favorably with their HER activity. As BN doped $C_{60}$ exhibits inherent potential to replace Pt-based electrocatalysts, meticulous manipulation of its bandgap through defect engineering (such as altering the BN arrangement) can prove instrumental in fabricating efficient non-metal photo- and electro-catalysts for the HER.\\
The frequency-dependent complex dielectric constant has been determined through initial principles calculations utilizing Density Functional Theory (DFT) within the random phase approximation (RPA)\cite{ahuja}. This computation is employed to assess and characterize the optical attributes of materials. In this context, $\epsilon_{1}$ and $\epsilon_{2}$ denote the real and imaginary components of the complex dielectric functions, and can be computed using equation ~\ref{eq:optical}, respectively\cite{mahida,mahida1,singh,ahuja1}. The intricate frequency-dependent dielectric constant is an instrumental parameter in discerning various optical characteristics of materials, including optical absorption, electron energy loss spectra, refractivity, extinction coefficient, reflectivity, and transmittance, etc.\cite{khire,hadji}.
\begin{equation}\label{eq:optical}
\begin{aligned}
	\epsilon_1(\omega)=1+\frac{2}{\pi}p\int_0^\infty \frac{\omega' \epsilon_2(\omega')d\omega'}{(\omega'^2-\omega^2)} \\
	\epsilon_2(\omega)=\frac{4\pi^2 e^2}{m^2 \omega^2} \sum_{ij} \int_k \langle i |M_j \rangle^2 f_i(1-f_i)X\delta(E_{jk}-E_{ik}-\omega)d^3k
\end{aligned}
\end{equation}
Figure~\ref{Fig:real_img} visually presents both the real and imaginary segments of the dielectric constant for the monolayers $C_{60}$, $C_{58}B_{1}N_{1}$, and $C_{54}B_{3}N_{3}$. The real portion of the dielectric function conveys the electronic polarizability of the substance and can be deduced via the Clausius-Mossotti relation\cite{ahuja1}. This aspect reveals how the real component of the complex dielectric function offers insights into the electronic polarizability of the materials. The static dielectric function (optical dielectric constant) pertains to the real part of the complex dielectric function at zero photon energy. Conversely, the imaginary part of the complex dielectric function corresponds to the inter-band transition of electrons from the valence to the conduction band.\\ 
Notably, Figure~\ref{Fig:real_img}(a,b,c) depicts that the static dielectric constant is approximately 3.56, 3.53, and 3.35 eV for in-plane ($E||X$) and around 3.06, 3.10, and 3.07 eV for out-of-plane ($E\perp Z$) polarization. The data showcases an anisotropic behavior in optical properties, with the real dielectric function exhibiting comparatively more significant electronic polarization along the in-plane direction. A noteworthy negative value at 3.21 eV emerges in the $E||X$ orientation for the $C_{60}$. This points to the monolayer $C_{60}$ displaying a metallic characteristic at this specific photon energy. It is interesting to note that for the case of $C_{58}B_{1}N_{1}$, and $C_{54}B_{3}N_{3}$, we didn't found any negative value neither in ($E||X$) nor in ($E\perp Z$) polarization. Moreover, three significant peaks are situated at 2.60(2.30,2.31) eV, 4.05(4.30,4.05) eV, and 9.77(9.27,9.17) eV for the $E||X$ polarization, and at 2.50(2.35,2.55) eV, 4.16(4.16,4.16) eV, and 10.02(10.42,8.61) eV for the $E\perp Z$ polarization direction for $C_{60}$($C_{58}B_{1}N_{1}$, and $C_{54}B_{3}N_{3}$). The maximum electronic polarizability is identified within the energy range of 2.60(2.30,2.31) eV $E||X$) and 2.50(2.35,2.55) ($E\perp Z$) for the $C_{60}$($C_{58}B_{1}N_{1}$, and $C_{54}B_{3}N_{3}$) single-layer sheet.
Figure~\ref{Fig:real_img}(d,e,f) visualizes the imaginary part of the complex dielectric function for all three systems, which is closely linked to interband transitions. Notably, the imaginary part's behavior in the low-frequency range demonstrates minimal responsiveness to electromagnetic radiation up to approximately 1.50(1.92,1,61) eV for parallel polarization and about 1.6(1.95,1.75) eV for perpendicular polarization for $C_{60}$($C_{58}B_{1}N_{1}$, and $C_{54}B_{3}N_{3}$). This aligns with the findings of the Density of States (DOS) analysis depicted in Fig.3(HSE results), indicating a lack of inter-band transitions from the Valence Band Maximum (VBM) to the Conduction Band Minimum (CBM) within the electronic bandgap.\\
\begin{figure*}[ht]
        \includegraphics[width=7.0in, height = 4.5in]{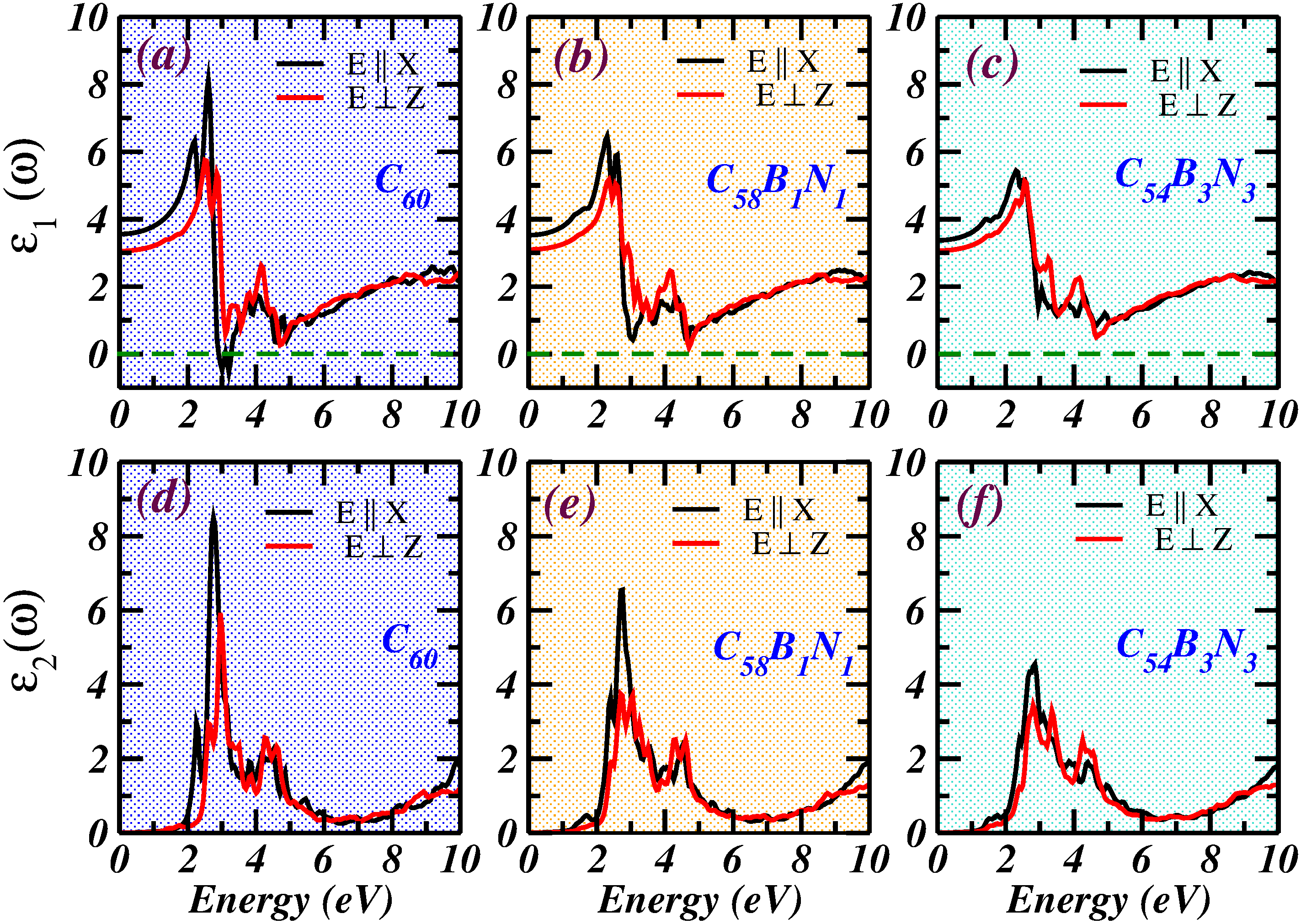}
	\caption{(a) Real ($\epsilon_{1}$) and (b) imaginary ($\epsilon_{2}$) diectric function of the single layer of $C_{60}$, $C_{58}B_{1}N_{1}$, and $C_{54}B_{3}N_{3}$.}\label{Fig:real_img}
\end{figure*}
Furthermore, Figure~\ref{Fig:abs_eels} portrays the computed frequency-dependent absorption coefficient averaged alover all the three distinct electric field orientations, a critical parameter for evaluating optical properties in optoelectronic applications. As an electromagnetic wave propagates through a medium over a unit distance, the degree of light intensity attenuation during propagation defines the light absorption coefficient. This coefficient shares a direct link with both the imaginary component of the dielectric function and the extinction coefficient, as indicated in Figure~\ref{Fig:real_img}(d,e,f). The absorption coefficient, denoted as $(\alpha(\omega))$, characterizes the depth to which light penetrates a given material. Lower light absorption translates to a diminished absorption coefficient for the material.
The first absorption peak for all the three systems emerges at energies between 1.5-2.0 eV, respectively. The uppermost absorption peak aligns with the energy of $\sim$12.25 eV, situating it within the UV region. Notably, the absorption coefficient remains inconspicuous within the lower energy range of $\sim$1.5 eV. Four significant peaks are observed for $E||X$, located at 10.03 eV, 11.76 eV, 13.10 eV, and 15.94 eV. In the $E\perp Z$ orientation, three notable peaks are evident, with the first at 7.84 eV, followed by peaks at 11.46 eV and 13.20 eV, displaying high values of $(\alpha(\omega))$. The considerable $\alpha(\omega)$ value of $1.07 x 10^{6} cm^{-1}$ suggests the potential application of the monolayers $C_{60}$, $C_{58}B_{1}N_{1}$, and $C_{54}B_{3}N_{3}$ as a UV absorber.\\
The energy loss function $L(\omega)$ is depicted in Figure~\ref{Fig:abs_eels}(b). Multiple distinct peaks are evident at approximately 3.82 eV, 5.11 eV, and 12.50 eV for the electric field aligned parallel to the single layer of all sheets. These peaks correspond to the $(\pi +\sigma)$( plasmon excitation. Conversely, there are no noticeable peaks at energy levels below 1.5 eV for either parallel or perpendicular polarizations. A few faint peaks are observed within the energy range of 1.5 eV to 3.5 eV, which are attributed to subtle resonances of incident light. For the proposed monolayers, the plasmonic peaks experience a shift toward higher energies and display increased sharpness (blueshift) when the material reflects frequencies of electromagnetic radiation below the plasma frequency. This phenomenon occurs because the electrons within the substance effectively shield the electric field of the radiation. Conversely, if the frequency of electromagnetic radiation surpasses the plasma frequency, it is transferred through the material when the electrons within it are unable to shield it.
\begin{figure}
  	\includegraphics[width=3in]{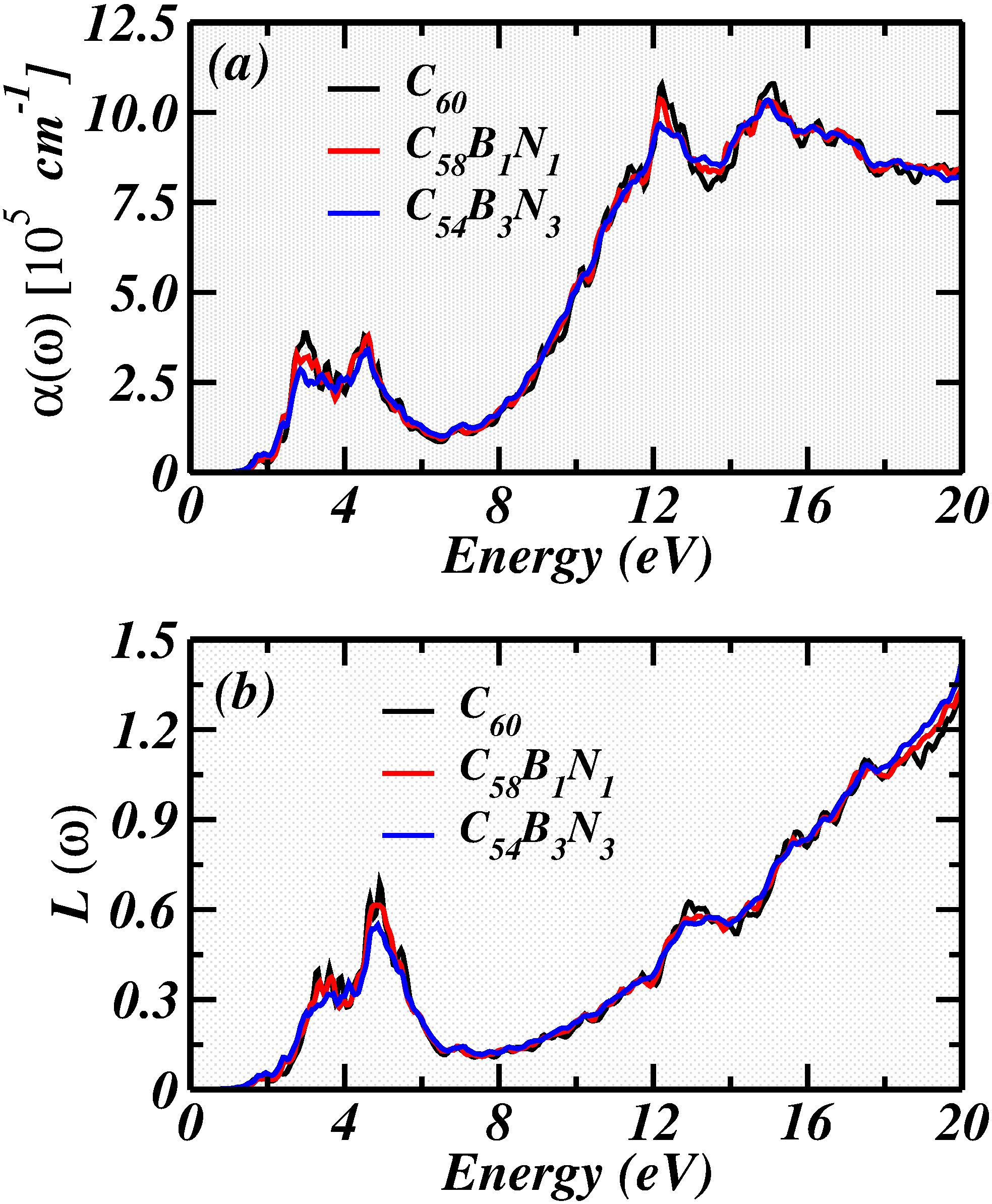}
	  \caption{(a) Absorption coefficient ($\alpha$) and (b) electron energy loss spectrum (L) of monolayers $C_{60}$, $C_{58}B_{1}N_{1}$, and $C_{54}B_{3}N_{3}$,respectively.}\label{Fig:abs_eels}
  \end{figure}
\section{Conclusions}
In this study, we explored the structural, chemical, electrical, and optical characteristics of pristine $C_{60}$ nanosheets. Additionally, we investigated the effects of BN doping in $C_{60}$ and compared the properties with the pristine one. Our approach employed quantum mechanical density functional theory (DFT) calculations. Our analysis determined that the calculated cohesion energy of all studied structures displayed significant negative values indicating their energetic stability. Specifically, both doped and pristine nanosheets were found to be energetically stable. Furthermore, our examination of global descriptor-DFT parameters revealed the order of chemical reactivity: $C_{58}B_{1}N_{1}$ > $C_{54}B_{3}N_{3}$ > $C_{60}$. This sequence suggests that all studied nanostructures possess a degree of chemical reactivity. Regarding their electronic properties, the analysis of the highest occupied molecular orbital-lowest unoccupied molecular orbital (HOMO-LUMO) gap and projected/partial density of states (PDOS) revealed that both pristine and BN-doped $C_{60}$ nanosheets are electrical semiconductors with bandgap ranges from 1.0-2.0 eV. Additionally, our investigation into the optical properties included the examination of electron absorption spectra and circular dichroism (CD). These analyses collectively provided evidence of the optical activity of all the studied nanosheets. In summary, this comprehensive study sheds light on the diverse properties of $C_{60}$ nanosheets and elucidates the influence of BN doping in $C_{60}$, enhancing our understanding of these intriguing nanomaterials. The present findings pave the way for the potential functionality of BN-doped $C_{60}$ to be utilized as efficient photovoltaic devices. The results presented here indicate the potential utility of BN-doped fullerene polymers in upcoming generations of 2D field-effect transistors. In forthcoming research, we intend to conduct ab initio calculations employing more dependable hybrid functionals with dopants other than BN to design semiconducting material efficient for catalysis and optoelectronic devices.

\section{Data availability}
The data required to reproduce these findings are available from the corresponding author upon reasonable request.

\section{Author Contributions}
VKY performed calculations, analyze the results and wrote the manuscript. 

\section{Acknowledgment}
VKY acknowledge the funding from University of Allahabad, Prayagraj. V.K.Y. also acknowledge National Supercomputing Mission (NSM) for providing computing resources of 'PARAM Shivay’ at Indian Institute of Technology (BHU), Varanasi, which are implemented by C-DAC and supported by the Ministry of Electronics and Information Technology (MeitY) and Department of Science and Technology (DST), Government of India.
\section{Keywords}
Density functional theory (DFT), $C_{60}$ fullerene network, BN doping, Optical properties, HER activity, Conductivity.


\end{document}